\documentclass[%
reprint,
superscriptaddress,
showpacs,preprintnumbers,
 amsmath,amssymb,
 aps,
longbibliography,
 prx,
floatfix,
]{revtex4-1}

\usepackage{float}
\usepackage{xspace}
\usepackage{graphicx}% Include figure files
\usepackage{dcolumn}% Align table columns on decimal point
\usepackage{bm}% bold math
\usepackage{color}
\usepackage{soul}
\usepackage{multirow}
\usepackage{hyperref}
\hypersetup{
    unicode=false,          % non-Latin characters in Acrobat’s bookmarks
    pdftoolbar=true,        % show Acrobat’s toolbar?
    pdfmenubar=true,        % show Acrobat’s menu?
    pdffitwindow=false,     % window fit to page when opened
    pdfstartview={FitH},    % fits the width of the page to the window
    pdftitle={Phonon anomalies in FeS},    % title
    pdfnewwindow=true,      % links in new window
    colorlinks=true,       % false: boxed links; true: colored links
    linkcolor=blue,          % color of internal links (change box color with linkbordercolor)
    citecolor=blue,        % color of links to bibliography
    filecolor=blue,      % color of file links
    urlcolor=blue,       % color of external links
    plainpages=false,        % hat Einfluss auf die Seitennummerierung
}

\newcommand{\TC}{\ensuremath{T_\mathrm{C}}\xspace}

\newcommand{\CXT}{\mbox{CrXTe$_3$}\xspace}
\newcommand{\CST}{\mbox{CrSiTe$_3$}\xspace}
\newcommand{\CGT}{\mbox{CrGeTe$_3$}\xspace}

\newcommand{\Ag}{\texorpdfstring{\ensuremath{A_{g}}\xspace}{Ag}}

\newcommand{\Eg}{\texorpdfstring{\ensuremath{E_{g}}\xspace}{Eg}}
\newcommand{\Au}{\texorpdfstring{\ensuremath{A_{u}}\xspace}{Ag}}
\newcommand{\Eu}{\texorpdfstring{\ensuremath{E_{u}}\xspace}{Eu}}

\newcommand{\wn}{\ensuremath{\rm cm^{-1}}\xspace}
\begin{document}

%%%%%%%%%%%%%%%%%%%%%%%%%%%%%%%%%%%%
%\begin{CJK*}{GBK}{}%%%%%%%%%%%%%%%%%
%%%%%%%%%%%%%%%%%%%%%%%%%%%%%%%%%%%%

\title{Evidence of spin-phonon coupling in CrSiTe$_3$}
\date{\today}
\author{A.~Milosavljevi\'{c}}
\affiliation{Center for Solid State Physics and New Materials, Institute of Physics Belgrade, University of Belgrade, Pregrevica 118, 11080 Belgrade, Serbia}
\author{A.~\v{S}olaji\'{c}}
\affiliation{Center for Solid State Physics and New Materials, Institute of Physics Belgrade, University of Belgrade, Pregrevica 118, 11080 Belgrade, Serbia}

\author{J.~Pe\v{s}i\'{c}}
\affiliation{Center for Solid State Physics and New Materials, Institute of Physics Belgrade, University of Belgrade, Pregrevica 118, 11080 Belgrade, Serbia}
\author{Yu Liu}
\affiliation{Condensed Matter Physics and Materials Science Department, Brookhaven National Laboratory, Upton, NY 11973-5000, USA}
\author{C.~Petrovic}
\affiliation{Condensed Matter Physics and Materials Science Department, Brookhaven National Laboratory, Upton, NY 11973-5000, USA}
\author{N.~Lazarevi\'{c}}
\affiliation{Center for Solid State Physics and New Materials, Institute of Physics Belgrade, University of Belgrade, Pregrevica 118, 11080 Belgrade, Serbia}
\author{Z.V.~Popovi\'{c}}
\affiliation{Center for Solid State Physics and New Materials, Institute of Physics Belgrade, University of Belgrade, Pregrevica 118, 11080 Belgrade, Serbia}
\affiliation{Serbian Academy of Sciences and Arts, Knez Mihailova 35, 11000 Belgrade, Serbia}

%%%%%%%%%%%%%%%%%%%%%%%%%%%%%%%%%%%%%%%%%%%%%%%%%%%%%%%%%%%%%%%%%%%%%%%%%%%%
\begin{abstract}
We present the Raman scattering results on layered 2D semiconducting ferromagnetic compound \CST. Four Raman active modes, predicted by symmetry, have been observed and assigned. The experimental results are supported by DFT calculations. The self-energies of the $\Ag^3$ and the $\Eg^3$ symmetry modes exhibit unconventional temperature evolution around 180\,K. In addition, the doubly degenerate $\Eg^3$ mode shows clear change of asymmetry in the same temperature region. The observed behaviour is consistent with the presence of the previously reported short-range magnetic order and the strong spin-phonon coupling.
\end{abstract}
%%%%%%%%%%%%%%%%%%%%%%%%%%%%%%%%%%%%%%%%%%%%%%%%%%%%%%%%%%%%%%%%%%%%%%%%%%%%%%%%%%%%%%%%%%%%%%%
\pacs{%
%78.30.-j, %Infrared and Raman spectra
%74.72.-h, %cuprate superconductors
%74.70.Xa, %pnictides and chalcogenides
%75.10.Jm, %quantized spin models including frustration
%74.20.Mn, %nonconventional mechanisms
%74.25.nd %Raman and optical spectroscopy (of superconductors)
}
\maketitle

%%%%%%%%%%%%%%%%%%%%%%%%%
%\end{CJK*}%%%%%%%%%%%%%%
%%%%%%%%%%%%%%%%%%%%%%%%%

%%%%%%%%%%%%%%%%%%%%%%%%%%%%%%%%%%%%%%%%%%%%%%%%%%%%%%%%%%%%%%%%%%%%%%%%%%%%%%%%%%%%%%%%%%%%

\section{Introduction}

Trichalcogenides \CXT (X = Si, Ge) belong to a rare class of quasi - 2D semiconducting materials with a ferromagnetic order, band gap of 0.4 eV for Si and 0.7 eV for Ge compound, and Curie temperatures (\TC) of 32  and 61\,K respectively \cite{doi:10.1063/1.4914134, 1347-4065-55-3-033001,LEINEWEBER1997145, OUVRARD198827, Siberchicot1996, 0295-5075-29-3-011}. Because of their layered structure, due to the van der Waals (vdW) bonding, they can be exfoliated to mono and few-layer nanosheets, which, together with their semiconducting and magnetic properties makes an ideal combination for applications in the optoelectronics and nano-spintronics \cite{PhysRevB.91.235425, Novoselov666, Wang2012, Gong2017, Huang2017}. This was further supported by the observation of giant resistivity modulation of \CGT-based devices \cite{2053-1583-4-2-024009}.

From the X-ray diffraction study \cite{doi:10.1063/1.4914134}, it was revealed that \CST crystals are twined along $c$ - axes, the thermal expansion is negative at low temperatures and thermal conductivity shows strong magnon-phonon scattering effects. A very small single ion anisotropy favoring magnetic order along $c$ - axes and spin waves were found in \CST  by elastic and inelastic neutron scattering \cite{PhysRevB.92.144404}. Spin wave measurements suggest the absence of three dimensional correlations above \TC, whereas in-plane dynamic correlations are present up to 300\,K.
First-principles calculations suggested a possibility of graphene-like mechanical exfoliation for \CXT (X = Si, Ge) single crystals with conserved semiconducting and ferromagnetic properties \cite{C4TC01193G}.
The exfoliation of \CST bulk to a mono and few-layer 2D crystals onto Si/SiO\textsubscript{2} substrate has been achieved \cite{C5TC03463A} with the resistivity between 80\,K and 120\,K,  depending on the number of the layers. Critical exponents for \CST were also determined from theoretical analysis \cite{Liu2017}. 

Spin-phonon coupling in \CGT was investigated by Raman scattering experiments \cite{2053-1583-3-2-025035}. Splitting of the two lowest energy \Eg modes in the ferromagnetic phase has been observed and ascribed to the time reversal symmetry breaking by the spin ordering. Further more the significant renormalisation of the three higher energy modes self-energies below \TC provided additional evidence for the spin-phonon coupling \cite{2053-1583-3-2-025035}.
The external pressure induced effect on lattice dynamics and magnetization in \CGT has also been studied \cite{doi:10.1063/1.5016568}.

Raman spectrum of \CST single crystal was reported in Ref. \cite{doi:10.1063/1.4914134}, where three Raman active modes have been observed. Similar results have also been presented in Ref. \cite{C5TC03463A} for an ultrathin nanosheets of \CST. Here, we report the Raman scattering study of \CST single crystals, with the main focus on phonon properties in the temperature range between 100 \,K and 300\,K. Our experimental results were qualitatively different from those previously reported \cite{doi:10.1063/1.4914134,C5TC03463A}, but consistent with the results obtained for \CGT \cite{2053-1583-3-2-025035,doi:10.1063/1.5016568}. Furthermore, our data revealed the pronounced asymmetry of the $\Eg^3$ mode, which is suppressed at higher temperatures. $\Ag^3$ and $\Eg^3$ symmetry modes exhibit non-anharmonic self-energy temperature dependance in the region around 180\,K, related to the strong spin-lattice interaction due to short range magnetic order \cite{doi:10.1063/1.4914134}. Energies and symmetries of the observed Raman active modes are in good agreement with theoretical calculations.

%%%%%%%%%%%%%%%%%%%%%%%%%%%%%%%%%%%%%%%%%%
\section{Experiment and numerical method}
\label{sec:exp_theo}

Single crystals of \CST and \CGT were grown as described previously \cite{PhysRevB.96.054406}. For a Raman scattering experiment, Tri Vista 557 spectrometer was used in the backscattering micro-Raman configuration with a 1800/1800/2400\,groves/mm diffraction grating combination. Coherent Verdi G solid state laser with 532 nm line was used as an excitation source. Direction of an incident (scattered) light coincides with a crystallographic  $c$ - axes. Right before being placed in the vacuum, the samples were cleaved in the air. All the measurements were performed in the high vacuum ($10^{-6}$ mbar)  using a  KONTI  CryoVac continuous Helium flow cryostat with 0.5 mm thick window. Laser beam focusing was achieved through the microscope objective with $\times$50 magnification. All the spectra were corrected for the Bose factor.

%%%%%%%%%%%%%%%%%%%%%%%%%%%%%%%%%%%%%%%%%%%%%%%%%%%%%%%%%%%%
\begin{table}[t]
\caption{Calculated and experimental crystallographic lattice parameters for \CST ($|a| = |b|$), bond lengths, interlayer distance ($d$) and van der Waals gap (vdW gap).}
\label{ref:Table1}
\begin{ruledtabular}
\centering
\resizebox{\linewidth}{!}{%
\begin{tabular}{c c c }
\CST & Calculations ($\mathrm{\mathring{A}}$) & Experiment ($\mathrm{\mathring{A}})$ \cite{PhysRevB.95.245212}  \\ [1mm] \cline{1-3}  \\[-0.5em]
$a$              & 6.87 & 6.76 \\[1mm] %\hline \\[-0.5em]
$c$              & 19.81& 20.67\\[1mm]
$\mathrm{Si-Si}$ & 2.27 & 2.27 \\[1mm]
$\mathrm{Si-Te}$ & 2.52 & 2.51 \\[1mm]
$\mathrm{Cr-Te}$ & 2.77 & 2.78 \\[1mm]
$d$              & 6.86 & 6.91 \\[1mm]
vdW gap          & 3.42 & 3.42 \\

\end{tabular}}
\end{ruledtabular}
\end{table}
%%%%%%%%%%%%%%%%%%%%%%%%%%%%%%%%%%%%%%%%%%%%%%%%%%%%%%%%%%%%%%%%%%%%%

Density functional theory calculations were performed in the Quantum Espresso (QE) software package  \cite{QE-2009}, using the PBE exchange-correlation functional \cite{PhysRevLett.77.3865}, PAW pseudopotentials \cite{PhysRevB.50.17953,PhysRevB.59.1758} and energy cutoff for wavefunctions and the charge density of 85 Ry and 425 Ry, respectively. For $k$-point sampling, the Monkhorst-Pack scheme was used, with $\Gamma$ centered  $8\times8\times8$ grid. Optimization of the atomic positions in the unit cell was performed until the interatomic forces were minimized down to $10^{-6}\mathrm{Ry / \mathring{A}}$. In order to obtain the parameters accurately, treatment of the vdW interactions was included using the Grimme-D2 correction \cite{doi:10.1002/jcc.20495}.
Phonon frequencies were calculated in $\Gamma$ point within the linear response method implemented in QE. Calculated crystallographic properties obtained by relaxing the structures are in good agreement with XRD measurements \cite{PhysRevB.95.245212}. Comparison between our, calculated, and experimental results is presented in Table \ref{ref:Table1}.

\section{Results and Discussion}
\label{sec:results}

\subsection{Polarization dependence}

%%%%%%%%%%%%%%%%%%%%%%%%%%%%%%%%%%%%%%%%%%%%%%%%%%%%%%%%%%%%%%%
\begin{figure}[h!]
  \centering
  \includegraphics[width=85mm]{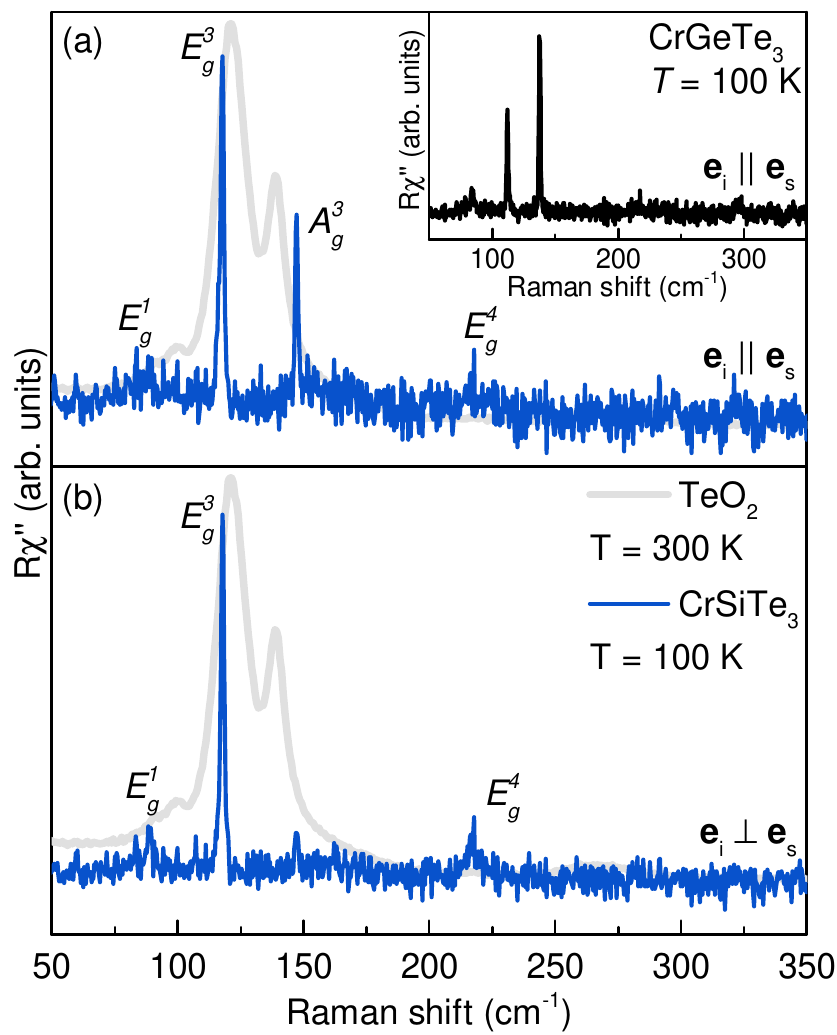}
  \caption{(Color  online) Raman spectra of \CST single crystal measured at 100\,K in (a) parallel and (b) cross polarisation configuration. Grey line represent TeO$_2$ spectrum measured and 300\,K. Inset: Raman spectrum of \CGT in parallel polarization configurations measured at 100\,K.}
 \label{fig:Figure1}
\end{figure}
%%%%%%%%%%%%%%%%%%%%%%%%%%%%%%%%%%%%%%%%%%%%%%%%%%%%%%%%%%

\CST crystallizes in the rhombohedral crystal structure, described with $R\overline{3}$ ($C_{3i}^2$) \cite{MARSH1988190}. Wyckoff positions of atoms, together with each site contribution to the phonons at $\Gamma$ point and corresponding Raman tensors are given in Table\,\ref{ref:Table2}. Phonon mode distribution obtained by factor group analysis for  $R\overline{3}$  space group is as follows:
\begin{eqnarray} \nonumber
\Gamma\textsubscript{Raman} = 5\Ag+5\Eg, \\ \nonumber
\Gamma\textsubscript{IR} = 4\Au+4\Eu, \\ \nonumber
\Gamma\textsubscript{Acoustic}=\Au+\Eu.
\end{eqnarray}
Since the plane of incidence is $ab$ plane, where  $|a| = |b|$ ($\measuredangle(a,b) = 120^{\circ}$), and direction of light propagation is along $c$ - axes, from the selection rules, it is possible to observe all Raman active modes, i.e. five \Ag modes and five doubly degenerate \Eg modes.
According to the Raman tensors presented in Table\,\ref{ref:Table2} \Ag symmetry modes are observable only in the parallel polarization configuration, whereas \Eg symmetry modes can be expected to appear for both, in parallel and cross polarization configuration.

%%%%%%%%%%%%%%%%%%%%%%%%%%%%%%%%%%%%%%%%%%%%%%%%%%%%%%%%%%%%%%
\begin{table}
\caption{Top panel: The type of atoms, Wyckoff positions, each site's contribution to the phonons in $\Gamma$ point and corresponding Raman tensors for $R\overline{3}$ space group of \CST. Bottom panel: Phonon symmetry, calculated optical phonon frequencies at 0\,K and experimental values for Raman (at 100\,K) and Infrared (at 110\,K) \cite{doi:10.1063/1.4914134} active phonons of \CST.}
\label{ref:Table2}
\begin{ruledtabular}
\centering
\resizebox{\linewidth}{!}{%
\begin{tabular}{c c c c c c}
\multicolumn{6}{c} {Space group  $R\overline{3}$ (No. 148)} \\

\cline{1-6} \\[-0.5em]

\multicolumn{3}{c} {Atoms (Wyckoff positions)} & \multicolumn{3}{c} {Irreducible representations} \\ [1mm] \cline{1-3} \cline{4-6}  \\[-0.3em]

\multicolumn{3}{c}{Cr, Si ($6c$)}
& \multicolumn{3}{c}{$A_{g} + E_g$+$A_{u} + E_u$} \\[1mm]

\multicolumn{3}{c}{Te ($18f$)}
& \multicolumn{3}{c}{3$A_{g} + 3E_g$+3$A_{u} + 3E_u$} \\[1mm]
\hline \\[-0.5em]

\multicolumn{6}{c}{Raman tensors} \\[1mm] \hline \\[-0.3em]

\multicolumn{2}{c}{$
A_{g} = \begin{pmatrix}
a&0&0\\
0&b&0\\
0&0&c\\
\end{pmatrix}
$}
&
\multicolumn{2}{c}{$
E^1_{g} = \begin{pmatrix}
c&d&e\\
d&-c&f\\
e&f&0\\
\end{pmatrix}
$}
&
\multicolumn{2}{c}{$E^2_{g} = \begin{pmatrix}
d&-c&-f\\
-c&-d&e\\
-f&e&0\\
\end{pmatrix}$}\\[5mm] \cline{1-6}  \\[-0.3em]

\multicolumn{3}{c}{Raman active} & \multicolumn{3}{c}{IR active \cite{doi:10.1063/1.4914134}} \\ [1mm] \cline{1-3} \cline{4-6} \\[-0.3em]

\multirow{ 2}{*}{Sym.} & Calc. & Exp. & \multirow{ 2}{*}{Sym.} & Calc. &  Exp.\\[1mm]
	 & (\wn) &(\wn) & &(\wn) & (\wn)\\[1mm] \cline{1-3} \cline{4-6}  \\[-0.5em]
$\Ag^1$ & 88.2  & -    & $\Au^1$ & 91.8 & 91.0 \\[1mm]
$\Eg^1$ & 93.5  & 88.9 & $\Eu^1$ & 93.7 &  -   \\[1mm]
$\Eg^2$ & 96.9  &  -   & $\Au^2$ & 116.8&  -   \\[1mm]
$\Eg^3$ & 118.3 &118.2 & $\Eu^2$ & 117.1&  -   \\[1mm]
$\Ag^2$ & 122.0 &  -   & $\Au^3$ & 202.4&  -   \\[1mm]
$\Ag^3$ & 148.0 &147.4 & $\Eu^3$ & 206.2& 207.9\\[1mm]
$\Ag^4$ & 208.7 &  -   & $\Au^4$ & 243.7&  -   \\[1mm]
$\Eg^4$ & 219.5 &217.2 & $\Eu^4$ & 365.8& 370.4\\[1mm]
$\Eg^5$ & 357.4 & -  &  &  \\[1mm]
$\Ag^5$ & 508.8  & -  &  & \\[1mm]
\end{tabular}}
\end{ruledtabular}
\end{table}
%%%%%%%%%%%%%%%%%%%%%%%%%%%%%%%%%%%%%%%%%%%%%%%%%%%%%%%%%%%%

The Raman spectra of \CST for two main linear polarization configurations, at 100\,K are shown in Figure\,\ref{fig:Figure1}. Four peaks can be observed in the spectra at energies of 88.9 \wn, 118.2 \wn, 147.4 \wn and 217.2 \wn. Since only the peak at 147.4 \wn vanishes in the cross polarization configuration, it corresponds to \Ag symmetry mode. Other three modes appear in both parallel and cross polarization configuration, thereby can be assigned as \Eg symmetry modes [Fig.\,\ref{fig:Figure1}].

In order to exclude the possibility that any of the observed features originate from the TeO$_2$  \cite{2053-1583-3-2-025035, PhysRevB.89.224301}, its Raman spectrum is also presented in a Fig.\,\ref{fig:Figure1}. It can be noted that no TeO$_2$ contribution is present in our \CST data. Furthermore, the observed \CST Raman spectra are also consistent with the \CGT Raman spectra (see Inset of a Fig.\,\ref{fig:Figure1}), isostructural to \CST. Five Raman active modes have been observed for \CGT, two \Ag modes at 137.9 \wn and 296.6 \wn, and three \Eg modes at 83.5 \wn, 112.2 \wn and 217.5 \wn, in agreement with the previously published data \cite{2053-1583-3-2-025035,doi:10.1063/1.5016568}. The main difference in the spectra of \CST and \CGT arises from the change of mass and lattice parameters effects that are causing the peaks to shift. 

Calculated and observed Raman active phonon energies are compiled in Table\,\ref{ref:Table2} together with the experimental energies of the 
infrared (IR) active phonons cite{doi:10.1063/1.4914134}, and are found to be in good agreement. Displacement patterns of \Ag and \Eg symmetry modes are presented in the Fig.\,\ref{ref:FigureA1} of the Appendix.

\subsection{Temperature dependence}
\label{sec:T}

%%%%%%%%%%%%%%%%%%%%%%%%%%%%%%%%%%%%%%%%%%%%%%%%%%%%%%%%%%%%%%%
\begin{figure}[h!]
  \centering
  \includegraphics[width=85mm]{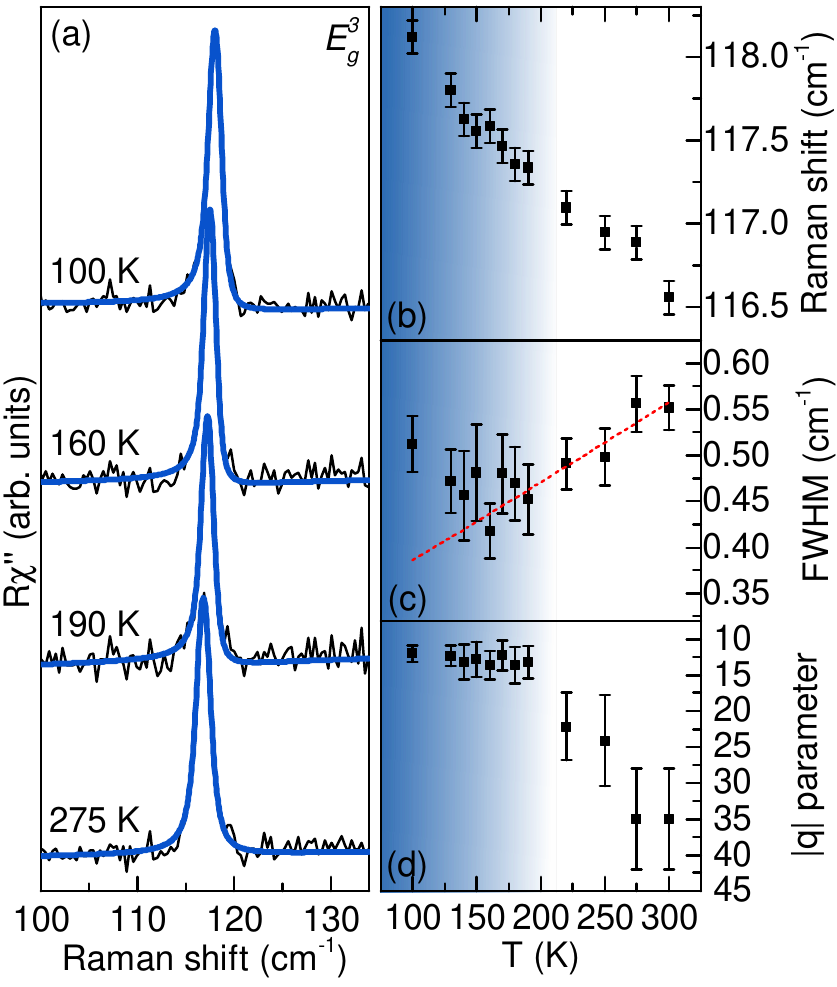}
  \caption{(Color  online) (a) The $\Eg^3$ mode Raman spectra of \CST at four different temperatures measured in cross polarization configuration. Blue lines represent line shape obtained as a convolution of Fano line shape and Gaussian and calculated to fit the experimetal data. Temperature  dependence of (b) energy, (c) linewidth and (d) Fano parameter $q$ of the $\Eg^3$ mode. Red dashed line represent standard anharmonic behaviour \cite{opacic,Baum2018_PRB97_054306}. All the parameters show discontinuity at about 180\,K.}
 \label{ref:Figure2}
\end{figure}
%%%%%%%%%%%%%%%%%%%%%%%%%%%%%%%%%%%%%%%%%%%%%%%%%%%%%%%%%%

After proper assignment of all the observed \CST Raman active modes we proceeded with temperature evolution of their properties, focusing on the most prominent ones, $\Eg^3$ and $\Ag^3$.
Figure\,\ref{ref:Figure2}\,(a) shows spectral region of doubly degenerate $\Eg^3$ mode at energy of 118.2 \wn, at four different temperatures. Closer inspection of the 100\,K spectra revealed clear asymmetry of the peak at the low energy side. The presence of defects may result in the appearance of the mode asymmetry \cite{PhysRevB.87.144305}, however, they would also contribute to the mode linewidth and, possibly, appearance of the phonons from the edge of the Brillouin zone in the Raman spectra \cite{Baum2018_PRB97_054306}. Very narrow lines and absence of the additional features in the Raman spectra of \CST do not support this scenario. The asymmetry may also arise when the phonon is coupled to a continuum \cite{Lazarevic2010_PRB81_144302}. Such coupling of the $\Eg^3$ phonon mode would result in a line shape given by the convolution of a Fano function and a Gaussian, the latter representing the resolution of the spectrometer \cite{Baum2018_PRB97_054306}.
Comparison between the Fano line shape convoluted with a Gaussian, the Voigt line shape and the experimental data at 100\,K is presented in Figure\,\ref{ref:FigureA2} of the Appendix with the former yielding better agreement to the experimental data.  Furthermore, it fully captures the $\Eg^3$ mode line shape at all temperatures under investigation [Fig.\,\ref{ref:Figure2}\,(a), Fig.\,\ref{ref:FigureA3}].

Upon cooling the sample, the $\Eg^3$ mode energy hardens [Fig.\,\ref{ref:Figure2}\,(b)] with a small discontinuity in the temperature range around 180\,K. Down to the same temperature, the linewidth monotonically narrows in line with the standard anharmonic behaviour (red dashed line in Fig.\,\ref{ref:Figure2}\,(c)). By further cooling, the linewidth increased, deviating from the expected anharmonic tendency. This indicates activation of an additional scattering mechanism, e.g. spin-phonon interaction. Fig.\,\ref{ref:Figure2}\,(d) shows the evolution of the Fano parameter, $|q|$. Whereas in the region below 180\,K, it increases slightly but continuously, at the higher temperatures it promptly goes to lower values and the mode recovers symmetric line shape. We belive that the observed behaviour of the $\Eg^3$ mode can be traced back to the short range magnetic correlations which, according to Ref. \onlinecite{doi:10.1063/1.4914134}, persist  up to 150\,K and the strong spin-phonon coupling in \CST. Similar behaviour of energy and linewidth, which differs from conventional anharmonic, as well as \Eg mode Fano-type lineshape, was recently reported in $\alpha$-RuCl\textsubscript{3} and was interpreted as a consequence of the spin-phonon interaction \cite{PhysRevLett.114.147201}.

%%%%%%%%%%%%%%%%%%%%%%%%%%%%%%%%%%%%%%%%%%%%%%%%%%%%%%%%%%%%%%%
\begin{figure}[t]
  \centering
  \includegraphics[width=85mm]{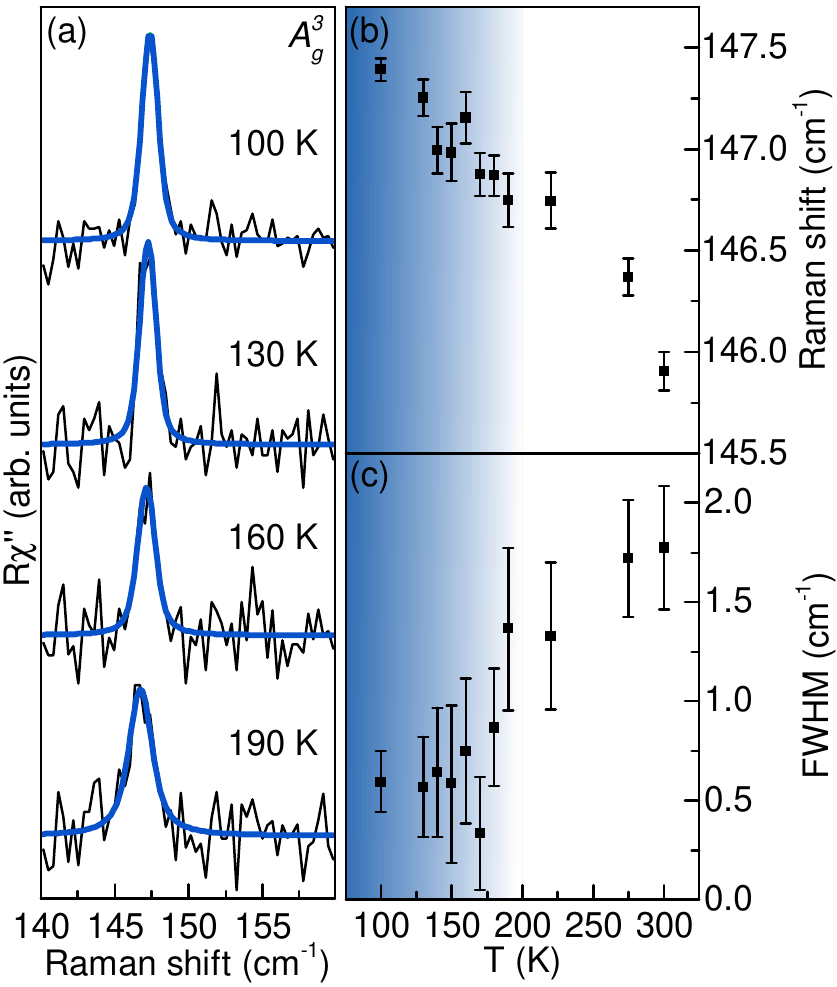}
  \caption{(Color  online) (a) The $\Ag^3$ mode Raman spectra of \CST at four different temperatures measured in parallel polarization configuration. Blue lines represent Voigt line shape. Energy (b) and (c) linewidth temperature  temperature dependence of The $\Ag^3$ mode.}
 \label{ref:Figure3}
\end{figure}
%%%%%%%%%%%%%%%%%%%%%%%%%%%%%%%%%%%%%%%%%%%%%%%%%%%%%%%%%%

Unlike the $\Eg^3$ mode, no pronounced asymmetry was observed for the  $\Ag^3$ mode. As can be seen form Fig.\,\ref{ref:Figure3}\,(b) and (c) both energy and linewidth of the $\Ag^3$ mode showed similar discontinuity in the same region of temperatures as the $\Eg^3$ mode, most likely due to the spin-phonon coupling.

\section{Conclusion}
\label{sec:conclusion}

The lattice dynamics of \CST, compound isostrucural to \CGT, is presented. An \Ag, and three \Eg modes were observed and assigned. The experimental results are well supported by theoretical calculations.  Temperature dependence of energies and linewidths of $\Ag^3$ and $\Eg^3$  modes, deviate from the conventional anharmonic model in the temperature range around 180\,K. In addition, $\Eg^3$  mode shows clear Fano resonance at lower temperatures. This can be related to the previously reported short-range magnetic correlations at temperatures up to 150\,K \cite{doi:10.1063/1.4914134} and strong spin-phonon coupling.

\section*{Acknowledgement}
The work was supported by the Serbian Ministry of Education, Science and Technological Development under Projects III45018 and OI171005. DFT calculations were performed using computational resources at Johannes Kepler University, Linz, Austria. Work at Brookhaven is supported by the U.S. DOE under Contract No. DESC0012704.

%\section*{Author contributions}

%A.B.,  conceived exp, performed exp., analyzed and discussed data, wrote paper \\
%A.M. conceived exp, performed exp., analyzed and discussed data, wrote paper \\
%N.L. conceived exp, performed exp., analyzed and discussed data, wrote paper \\
%M.M.R. calculated phonon dispersion and PDOS \\
%B.N. calculated MGPT \\
%M.M. performed exp., analyzed and discussed data \\
%Z.I.M. performed exp. \\
%M.S. performed exp. \\
%M.G.B. performed ellipsometric measurements\\
%N.S. performed exp. \\
%M.O. performed SQUID measurements \\
%A.W. synthesized and characterized the samples \\
%C.P. synthesized and characterized the samples \\
%Z.V.P. analyzed and discussed data, wrote paper \\
%R.H. conceived exp, analyzed and discussed data, wrote paper \\
%All authors commented on the manuscript.

%\bibliography{literature}

%\end{document}

%%%%%%%%%%%%%%%%%%%%%%%%%%%%%%%%%%%%%%%%%%%%%%%%

\clearpage
\begin{appendix}

\setcounter{figure}{0}
\renewcommand\thefigure{A\arabic{figure}}

\setcounter{table}{0}
\renewcommand\thetable{A\Roman{table}}

\section*{Eigenvectors of Raman active modes}

\begin{figure}[H]
  \centering
  \includegraphics[width=85mm]{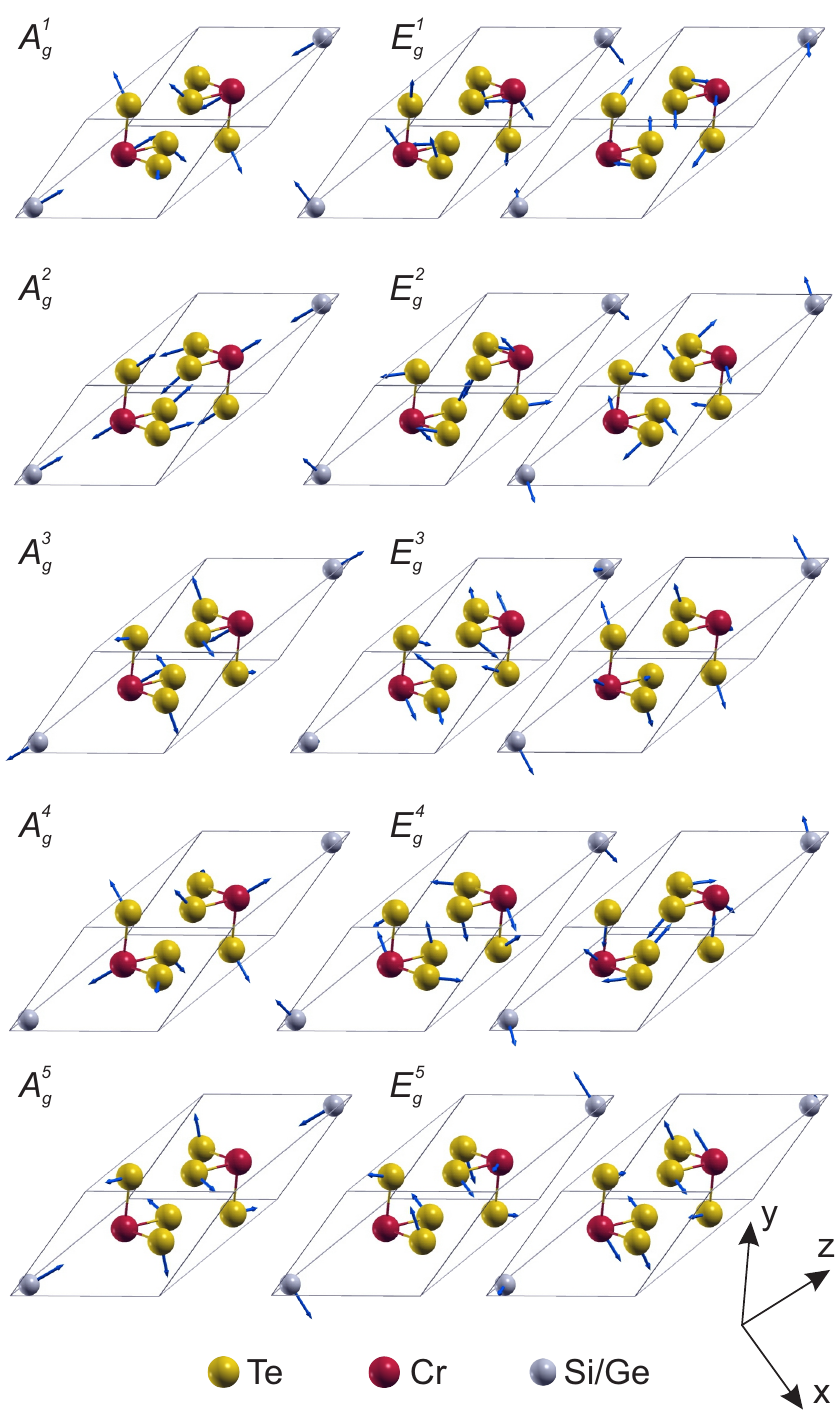}
  \caption{(Color  online) Unit cell of \CST single crystal (solid lines) with the displacement patterns of \Ag and \Eg symmetry modes. Arrow lengths are proportional to the square root of the inter-atomic
forces.}
 \label{ref:FigureA1}
\end{figure}

Figure\,\ref{ref:FigureA1} summarizes \Ag and \Eg symmetry modes displacement patterns of \CST single crystal ($R\overline{3}$  space group). Arrow lengths are proportional to the square root of the inter-atomic forces. 

\section*{Asymmetry of the $\Eg^3$ line}

\begin{figure}[H]
  \centering
  \includegraphics[width=85mm]{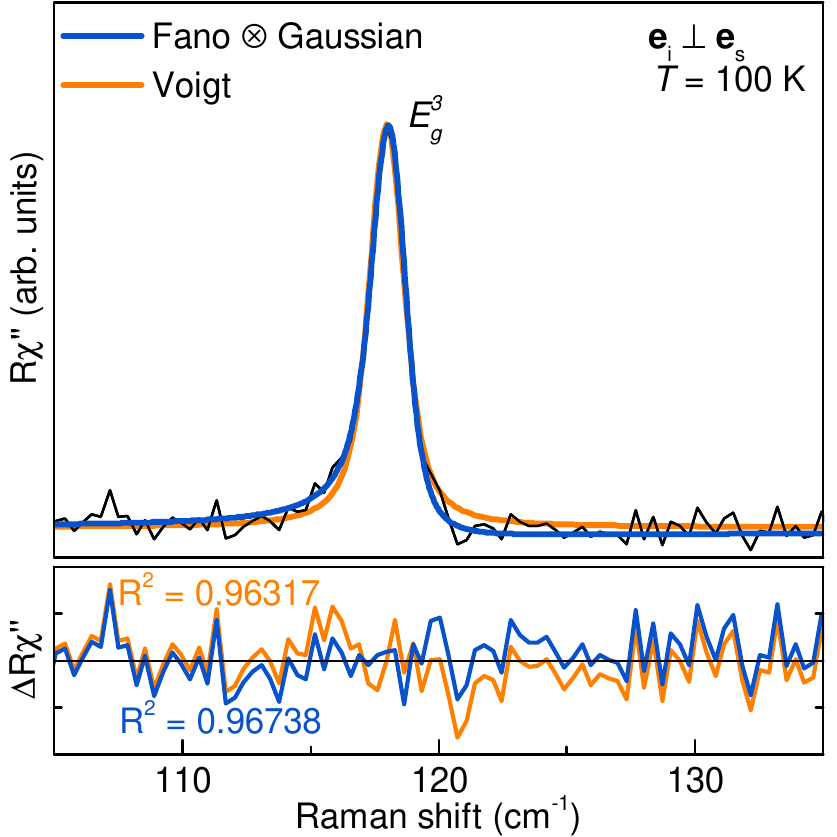}
  \caption{(Color  online) The analysis of the $\Eg^3$ asymmetry. Measured data are shown as the black line. The  blue solid line represents the line shape obtained as a convolution of  the Fano line shape and a Gaussian whereas the orange line represents a Voigt  line shape, both calculated to fit the experimental data. The Voigt profile  deviates from the experimental data at the peak flanks. }
 \label{ref:FigureA2}
\end{figure}

The  peak  at  118.2 \wn, that we assigned as $\Eg^3$ symmetry mode,
at  low  temperatures  shows  a significant  asymmetry  towards  lower  energies. Coupling  of  the phonon mode to a continuum may result in an  asymmetric lineshape described with Fano function. Due to the finite resolution of the spectrometer it has to be convoluted with a Gaussian ($\Gamma_G=1$ \wn). In Fig.\,\ref{ref:FigureA2} we present the comparison of the line obtained as a convolution of the Fano line shape and a Gaussian (blue line), and a Voigt line shape (orange line) fitted to the experimental data. Whereas the Voigt lineshape deviates at the peak flanks, excellent agreement has been achieved for convolution of Fano line shape and a Gaussian.

%%%%%%%%%%%%%%%%%%%%%%%%%%%%%%%%%%%%%%%%%%%%%%%%%%%%%%%%%%%%%%%
%%%%%%%%%%%%%%%%%%%%%%%%%%%%%%%%%%%%%%%%%%%%%%%%%%%%%%%%%%

\section*{$\Eg^3$ mode temperature dependance}

\begin{figure}[H]
  \centering
  \includegraphics[width=85mm]{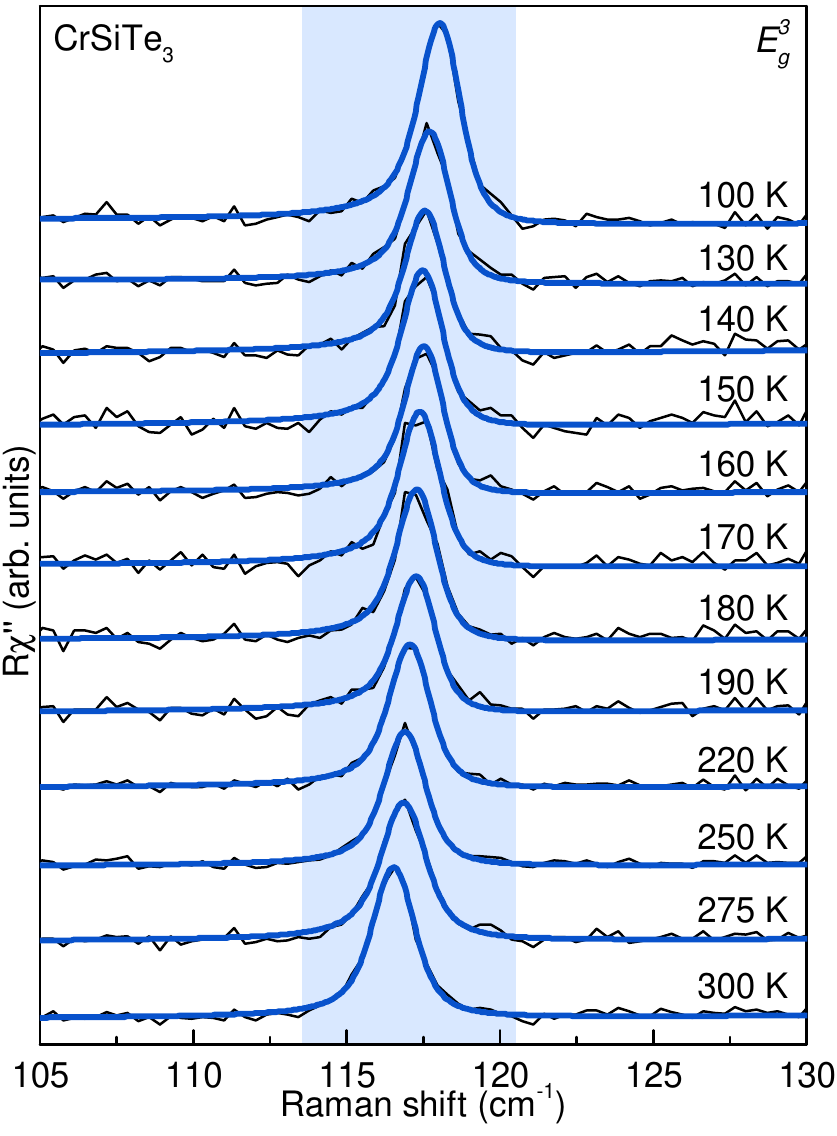}
  \caption{(Color  online) The $\Eg^3$ mode Raman spectra of \CST at all temperatures measured in cross polarization configuration. Blue lines represent calculated spectra obtained as convolution of Fano lineshape and Gaussian. }
 \label{ref:FigureA3}
\end{figure}

Figure \ref{ref:FigureA3} shows Raman spectra of \CST in the region of $\Eg^3$ mode in cross polarization configuration at various temperatures. Blue solid lines represent the convolution of Fano line shape and Gaussian fitted to the experimental data. The asymmetry is the most pronounced below 190 \,K. Above this temperature, the asymmetry is decreasing, and at high temperatures the peak recovers the fully symmetric line shape.

%%%%%%%%%%%%%%%%%%%%%%%%%%%%%%%%%%%%%%%%%%%%%%%%%%%%%%%%%%%%%%%
%%%%%%%%%%%%%%%%%%%%%%%%%%%%%%%%%%%%%%%%%%%%%%%%%%%%%%%%%%

\end{appendix}

\end{document}